\documentstyle[10pt,emulateapj]{article}

\input{epsf}

\journalid{}{}
\articleid{}{}
\begin{document}
\def\lax    {\ifmmode{_<\atop^{\sim}}\else{${_<\atop^{\sim}}$}\fi}
\def\gax    {\ifmmode{_>\atop^{\sim}}\else{${_>\atop^{\sim}}$}\fi}
\def\gtorder{\mathrel{\raise.3ex\hbox{$>$}\mkern-14mu
             \lower0.6ex\hbox{$\sim$}}}
\def\ltorder{\mathrel{\raise.3ex\hbox{$<$}\mkern-14mu
             \lower0.6ex\hbox{$\sim$}}}
 
\long\def\***#1{{\sc #1}}
 
\title{On the X-ray source luminosity distributions in the bulge and disk 
of M31: First results from {\em XMM-Newton} survey.}

\author{Sergey P.~Trudolyubov\altaffilmark{1}, 
Konstantin N.~Borozdin\altaffilmark{1}, William C.~Priedhorsky}
\affil{Los Alamos National Laboratory, Los Alamos, NM 87545}

\altaffiltext{1}{{\it Also:} Space Research Institute, 
Russian Academy of Sciences, Profsoyuznaya 84/32, Moscow 117810, Russia}

\author{Keith O.~Mason}
\affil{Mullard Space Science Laboratory, University College London, 
Holmbury St. Mary, Dorking, Surrey, U.K.}

\author{France A.~Cordova}
\affil{Department of Physics, University of California, 
Santa Barbara, CA 93106}

\begin{abstract}
\noindent We present luminosity distributions for the X-ray sources detected 
with {\em XMM-Newton} in the bulge and disk of the Andromeda Galaxy (M31). 
The disk is clearly lacking the brighter sources which dominate X-ray 
emission from the bulge.  This is the first convincing evidence for a 
difference between bulge and disk X-ray populations in M31. Our results are 
in good qualitative agreement with the luminosity distributions for low- and 
high-mass X-ray binaries recently obtained by Grimm et al.(2001) for our 
Galaxy. This confirms that X-ray population of the disk of M31 is dominated 
by fainter HMXB sources, while the bulge is populated with brighter LMXBs.
\end{abstract} 

\section{INTRODUCTION}
Multiple observations of external galaxies starting with {\it Einstein}
have shown convincingly that the X-ray emission of normal spirals similar 
to our own Milky Way is dominated by X-ray binaries (\cite{Fabb95}).
An X-ray binary contains either a neutron star or a black hole; it can be 
classified as low-mass (LMXB) or high-mass (HMXB) according to the spectral 
type of the companion star. It is well known that in our Galaxy most HMXBs 
lie in the Galactic plane, while the LMXBs have more spherical distribution
with conspicuous concentration towards the Galactic Center. LMXBs and HMXBs 
demonstrate very similar X-ray spectra in some of their states, therefore 
difficult to discriminate them in a single spectral observation. Recently 
Grimm, Gilfanov and Sunyaev (2001) demonstrated the difference between X-ray 
luminosity distributions of LMXBs and HMXBs in the Milky Way, providing a 
method for discriminating the two populations in other galaxies.

M31 (the Andromeda Galaxy) is of special importance for extragalactic 
astronomy because, at a distance of 760 kpc (van der Bergh 2000), it 
is the closest spiral galaxy to our own. M31 is believed in many 
respects to be similar to the Milky Way. By observing M31 we can sample 
hundreds of X-ray sources at a nearly uniform distance and less obscured 
by interstellar gas and dust than those in the Galaxy due to the favorable 
orientation of the M31. It is therefore quite interesting to look at the 
luminosity distributions for various populations of X-ray sources in M31 
and compare them with our Galaxy. However, earlier attempts to find the 
difference between luminosity distributions of X-ray sources in the bulge 
and disk of M31 were unsuccessful (\cite{TF91}).

In this Letter we report on the difference between the disk and bulge 
X-ray source populations of M31, as observed with {\em XMM-Newton} in the 
course of the most sensitive X-ray survey of a neighboring galaxy to date.

\section{OBSERVATIONS AND DATA ANALYSIS}
\begin{figure}
\vbox{
\hbox{
\begin{minipage}{8.5cm}
\epsfxsize=8.0cm
\epsffile{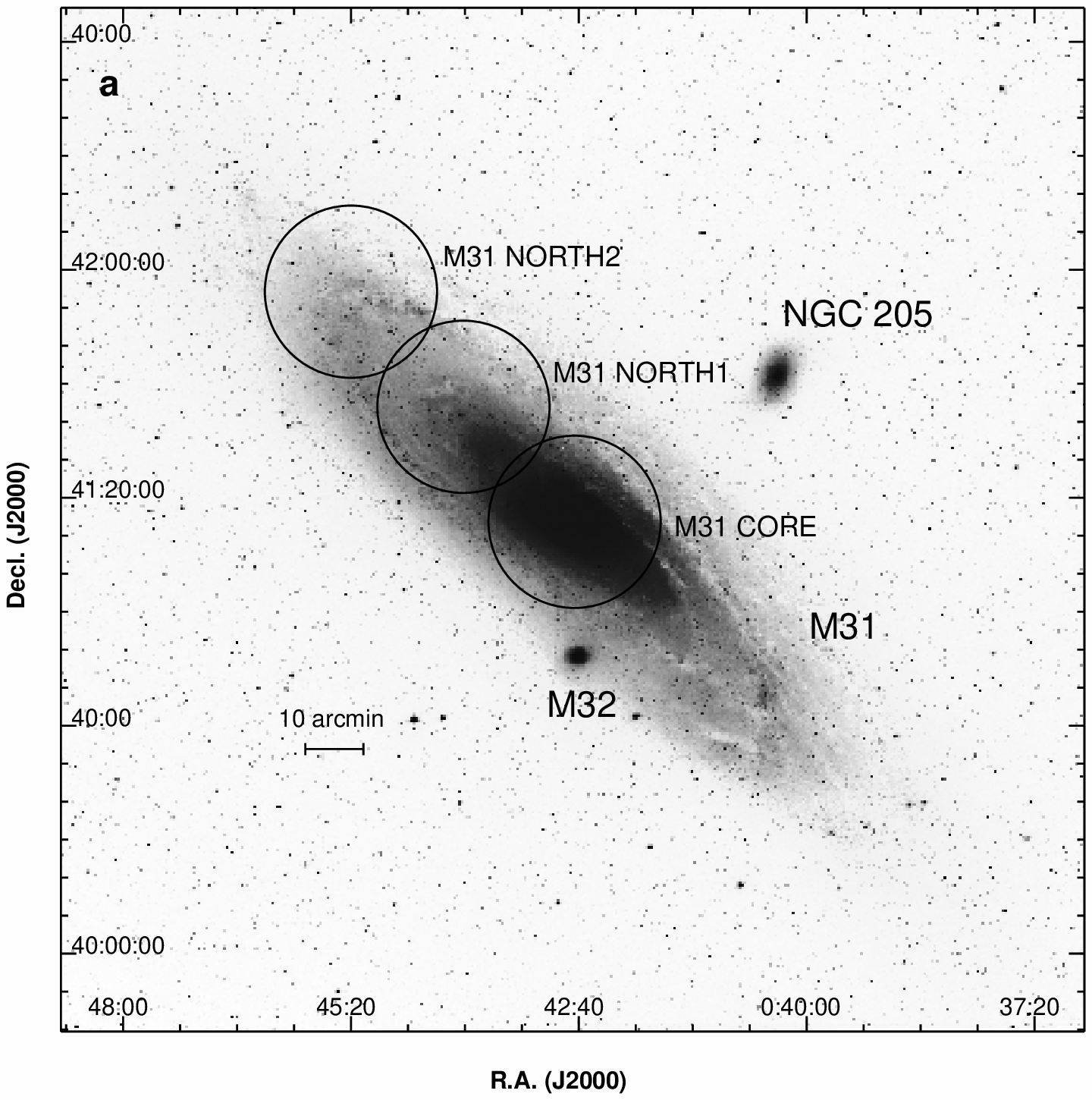}
\end{minipage}
\begin{minipage}{8.5cm}
\epsfxsize=8.0cm
\epsffile{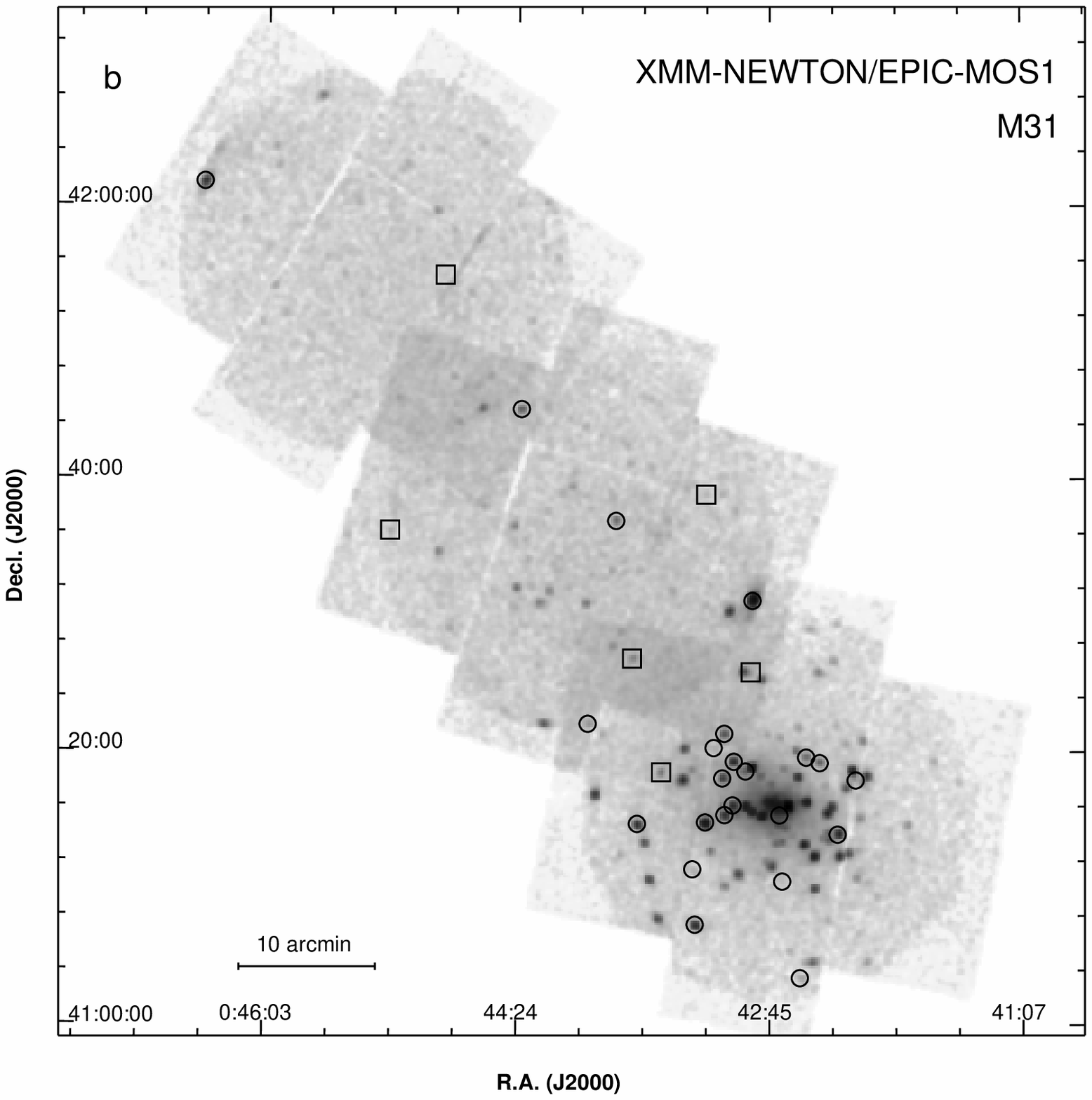}
\end{minipage}
}
\vspace{0.4cm}
\hbox{
\begin{minipage}{8.5cm}
\epsfxsize=8.0cm
\epsffile{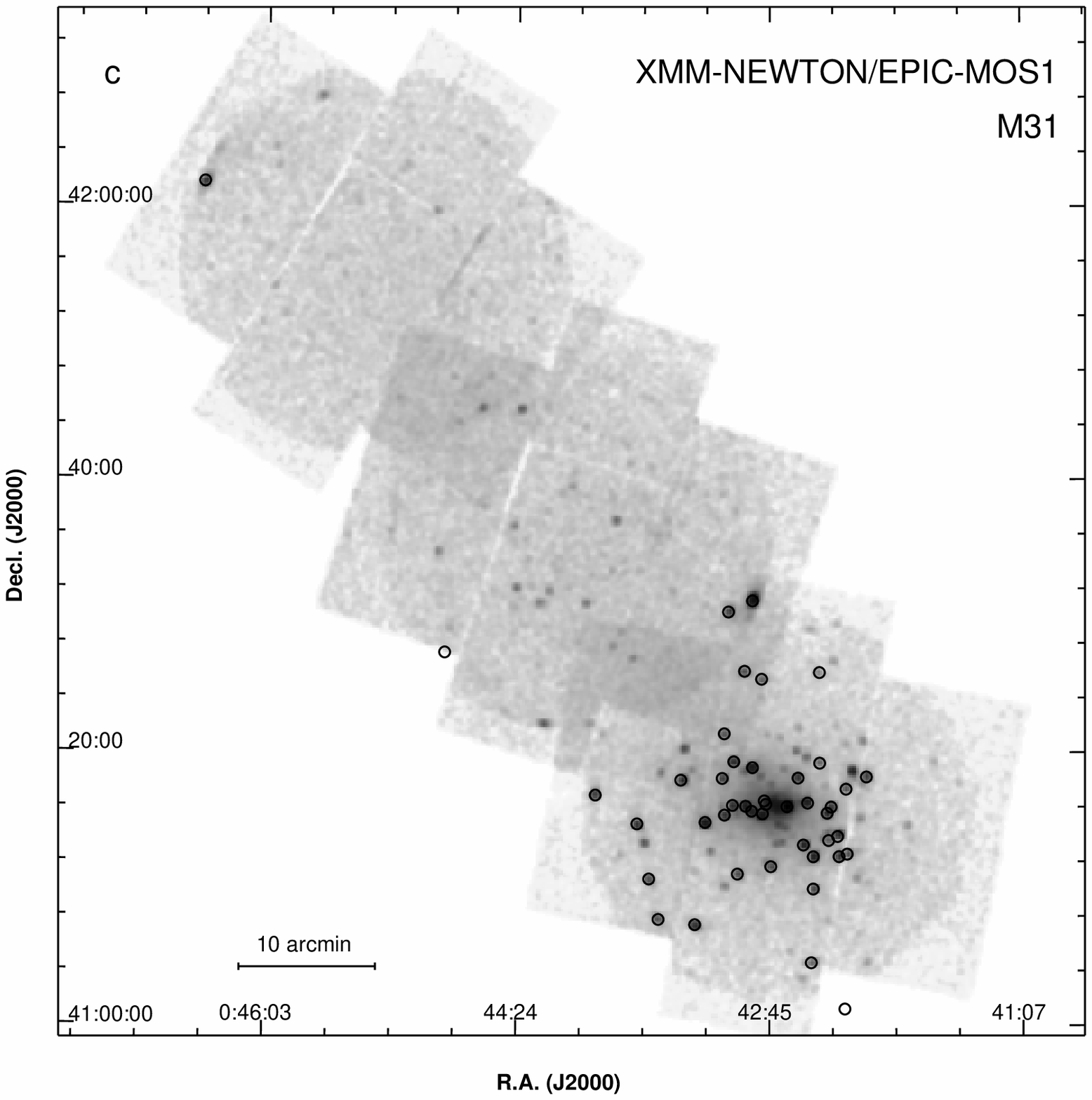}
\end{minipage}
\begin{minipage}{8.5cm}
\epsfxsize=8.0cm
\epsffile{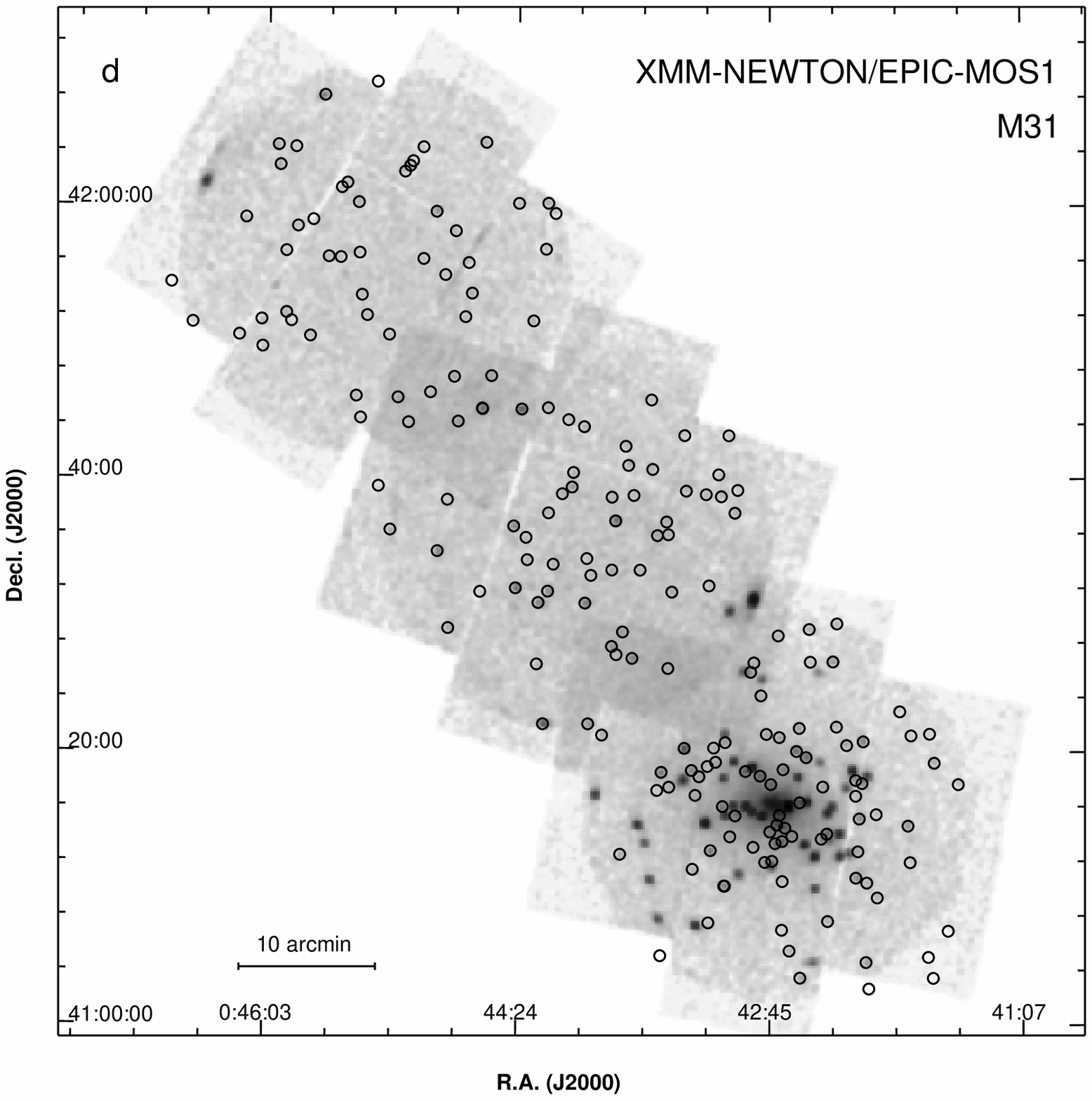}
\end{minipage}
}
}
\caption{Combined {\em XMM-Newton}/EPIC-MOS1 images of the M31 Core, 
North1 and North2 Fields. {\em Panel a} is an optical image of M31 
from the Digitized Sky Survey (DSS) with EPIC-MOS1 field of view shown 
by solid black circles for each of the observations. The X-ray emission 
intensity is shown by logarithmic grey scale (black is the maximum) 
in {\em Panels b,c,d} representing enlarged central (M31 Core; $r < 
15\arcmin$ and northern parts of the galaxy (North1 and North2 fields). 
Positions of identified globular cluster candidates (\cite{Bo87,Barmby01}) 
are shown with small circles in {\em Panel b}. Identified supernova remnants 
(\cite{Blair81,Magnier95}) are marked with small boxes in the same panel. In 
{\em Panel c} we circled all detected sources with luminosities greater than 
10$^{37}$ erg/s. Fainter sources are shown in {\em Panel d}. Some of the 
sources shown lie outside the EPIC-MOS1 sensitive area, but were detected 
with the EPIC-MOS2 or EPIC-PN detectors. It is evident that the brightest 
sources are concentrated towards the center of the galaxy, while fainter 
sources are distributed throughout the disk. A few bright sources in the 
disk regions (North1 and North2 fields) are identified with globular 
clusters, and hence belong to the non-disk subsystem of the galaxy. 
\label{image_general}}
\end{figure}

In the following analysis we use data from three {\em XMM-Newton} 
observations of the bulge and galactic disk of M31 (see Table \ref{obslog}, 
Fig. \ref{image_general}{\em a}). The first observation of the central part 
of M31 (M31 Core; $r < 15\arcmin$) was performed on June 25, 2000 
\cite{Shirey01,Osborne01}. Two regions of the northern part of the 
galactic disk ($r > 15\arcmin$) were observed with {\em XMM-Newton} on 
January 5 \cite{xmm_circ02} (M31 North1 Field) and January 26, 2002 (M31 
North2 Field) as a part of the Guaranteed Time Program (PI: K.O.~Mason)(Fig. 
\ref{image_general}). We use data from three European Photon Imaging Camera 
(EPIC) instruments: two EPIC MOS detectors (\cite{Turner01}) and the EPIC PN 
detector (\cite{Strueder01}). In all observations the EPIC instruments were 
operated in the {\em full window mode} ($30\arcmin$ diameter FOV) with the 
medium optical blocking filter. 
 
We reduced EPIC data with the {\em XMM-Newton} Science Analysis System (SAS 
v 5.2)\footnote{See http://xmm.vilspa.esa.es/user}. Images in celestial 
coordinates with a pixel size of 2$\arcsec$ have been accumulated in the 
$0.3 - 12$ keV energy band from both MOS and PN detectors. The resulting 
images for the EPIC-MOS1 instrument are shown in Figure 
\ref{image_general}{\em b,c,d}. Sources were detected with a program based 
on the wavelet decomposition algorithm, set at $4\sigma$ detection threshold. 
We expect errors in the source position determination to be dominated by 
residual systematic errors of order $2 - 5 \arcsec$.

We detected 230 point X-ray sources above a $4 \sigma$ level of significance. 
We cross-identified the detected X-ray objects with catalogs of globular 
cluster candidates and supernova remnants 
(\cite{Bo87,Barmby01,Blair81,Magnier95}), based on a positional coincidence of 
$< 6\arcsec$ (Fig. \ref{image_general}{\em b}). Four sources were identified 
with foreground objects and excluded from the analysis. Due to source 
confusion in the central part of the galaxy sources located within the 
central $1\arcmin$ were also excluded. 

To estimate energy fluxes and spectra, we used an extraction radius of $\sim 
20 - 60 \arcsec$ (depending on the distance of the source from the telescope 
axis) and subtracted as background the spectrum of adjacent source-free 
regions. We corrected the count rates for the vignetting of the XMM 
telescope, based on the Current Calibration Files provided with SAS. 
The EPIC count rates were converted into energy fluxes using analytical 
fits to the spectra of brighter sources. Detailed spectral analysis 
for individual sources will be presented elsewhere. For faint sources 
we estimated energy fluxes with Web-PIMMS
\footnote{See heasarc.gsfc.nasa.gov/Tools/w3pimms.html} assuming an absorbed 
simple power law spectral shape with photon index $\alpha = 1.5$ and an 
equivalent absorbing column of $7 \times 10^{20}$ cm$^{-2}$ (Galactic value
for the direction to M31).

\section{RESULTS AND DISCUSSION}
Altogether we detected 230 X-ray sources in the three fields 
of M31: 118 sources in the Core and 112 in the North1 and North2 fields. 
These include unidentified background and foreground sources, as discussed 
below. The properties of the individual sources are presented elsewhere 
(\cite{Shirey01,Osborne01,xmm_nn_catalog}). Here we discuss 
the source populations and their luminosity distributions. We have noticed 
that the brightest X-ray sources are concentrated towards the center of M31, 
while the fainter ones are distributed much more uniformly (Fig. 1{\em c,d}). 
Furthermore, we have mentioned that all remaining bright sources in North1 
and North2 fields can be identified with globular clusters (Fig.1{\em b}).
Our data prove, for the first time, that M31 resembles our Galaxy in the 
distribution of its X-ray source populations (i.e. LMXBs and HMXB) (see 
Fig.1 in (\cite{Grimm01})). 

\subsection{Luminosity distributions of X-ray sources}
\begin{figure}
\hbox{
\begin{minipage}{8.5cm}
\epsfxsize=8.0cm
\epsffile{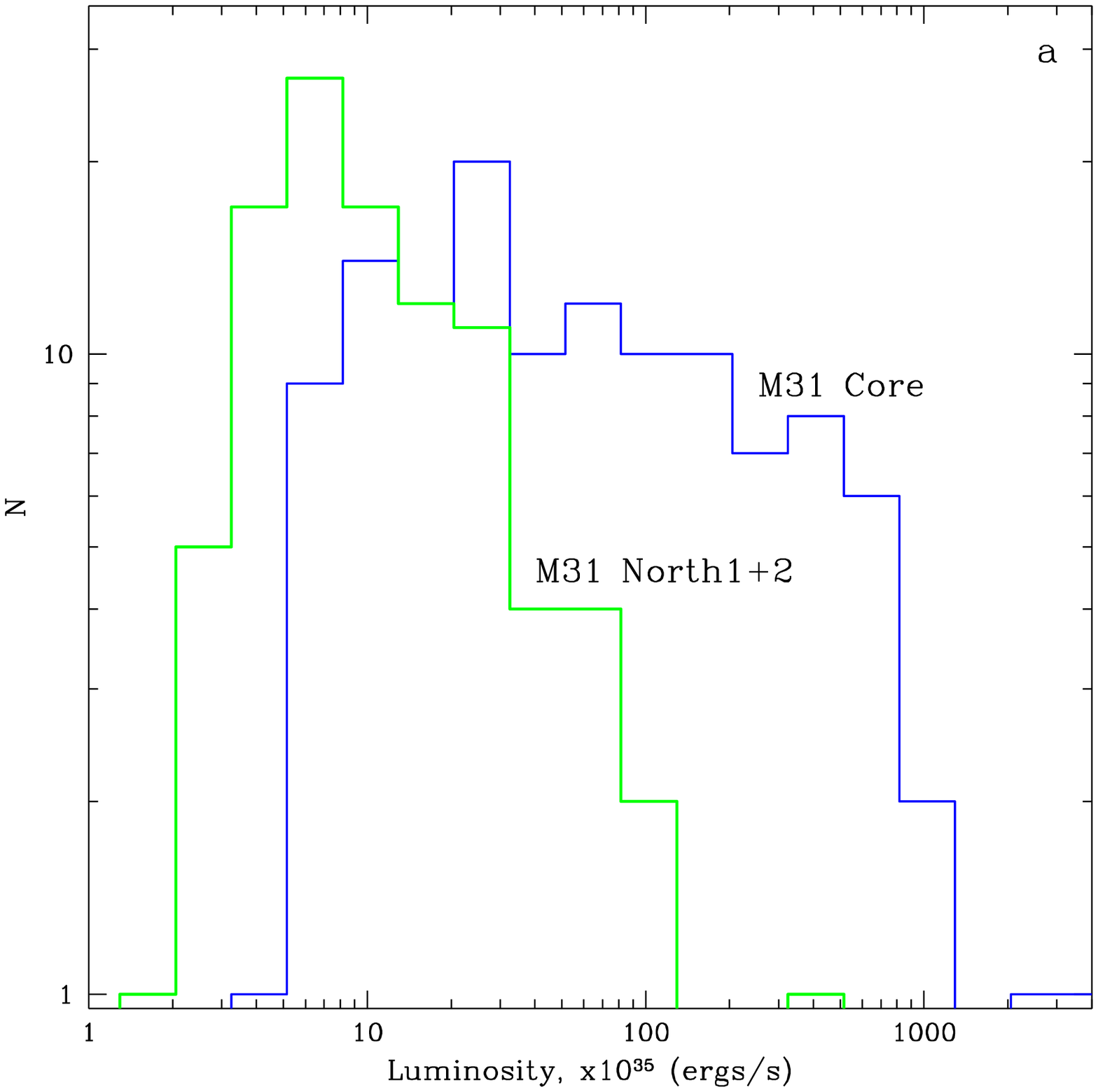}
\end{minipage}
\begin{minipage}{8.5cm}
\epsfxsize=8.0cm
\epsffile{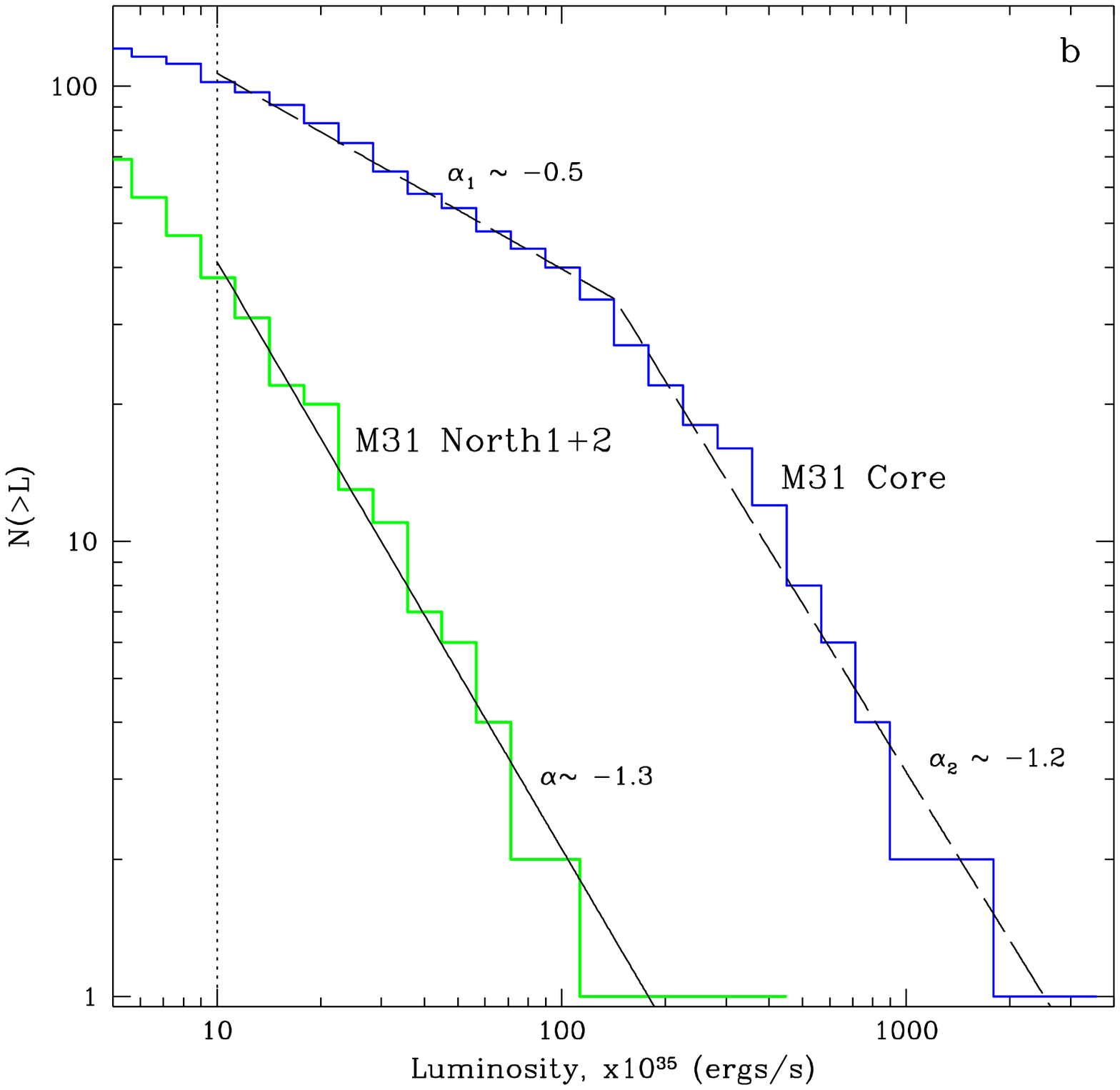}
\end{minipage}
}
\caption{{\em Panel a:} Differential luminosity distributions of the 
bulge (M31 Core) and disk (North1 and North2 regions combined) of M31. 
{\em Panel b:} Cumulative luminosity distributions for the same regions. 
A broken power law approximation to the luminosity distribution of the 
central part of the galaxy is shown with thin dashed line. A power law 
approximation to the luminosity distribution of the disk regions is shown 
with thin solid line. The indices of the power law fits to the luminosity 
distributions are also indicated. The vertical dotted line corresponds to 
the approximate completeness limit of our sample.
\label{lum_distr}}
\end{figure}

We have built differential and cumulative luminosity distributions of the 
detected X-ray sources for the central part of M31, and the combined North1 
and North2 fields assuming a distance of 760 kpc (Fig. \ref{lum_distr}). 
In order to minimize the contribution from non-disk X-ray population, the 
sources identified with globular clusters were excluded from the 
North1/North2 sample. Taking into account the sensitivity as a function of 
off-axis distance, the different exposure times, and the effect of diffuse 
X-ray emission (in the M31 Core field), we estimate a flux completeness 
limit of our sample as $\sim 10^{36}$ ergs s$^{-1}$ (indicated by the dotted 
line in Fig. \ref{lum_distr}{\em b}). 

Some of the objects in the M31 fields must be background and foreground 
sources lying outside M31 (i.e. background AGN, and Galactic K and M stars). 
Based on {\em Chandra} and {\em XMM-Newton} deep field results 
\cite{Giacconi01,Hasinger01}, one might expect to detect $< 10$ 
background AGN in each field with an apparent luminosity around $10^{36}$ 
ergs s$^{-1}$. The total number of unidentified foreground Galactic stars 
could be $\sim 5$ sources per field \cite{Shirey01}. Correction for such 
background/foreground objects would flatten the faint end of the cumulative 
luminosity distribution (below $\sim 3 \times 10^{36}$ ergs s$^{-1}$), but 
would not significantly modify the higher luminosity part of the distribution.

As it is clearly seen from both panels of Fig. \ref{lum_distr}, the 
population of the northern disk regions of M31 is dominated by relatively 
faint objects with luminosities below $\sim 10^{37}$ ergs s$^{-1}$. The 
overall shape of the cumulative luminosity distribution (Fig. 
\ref{lum_distr}{\em b}) can be presented with a simple power law function 
with an index of $-1.3 \pm 0.2$ \footnote{We fitted unbinned cumulative 
luminosity distributions using a maximum likelihood method.} for luminosities 
above $\sim 10^{36}$ ergs s$^{-1}$. There is some indication for a flattening 
of the distribution at lower luminosities (Fig. \ref{lum_distr}{\em b}), 
which can be attributed to the incompleteness of our sample at the faint end. 

In contrast to the northern disk regions, bright X-ray sources contribute
significantly to the population in the central region of M31 (Fig. 
\ref{lum_distr}{\em a}). The shape of both cumulative and 
differential luminosity distributions for the central region of M31 ({\em 
Blue histograms} in Fig. \ref{lum_distr}) indicate the presence of a brake 
at around $(1 - 2) \times 10^{37}$ ergs s$^{-1}$. We fit the unbinned 
cumulative distribution with a broken power law (Fig. 
\ref{lum_distr}{\em b}). For source luminosities between $10^{36}$ and 
$1.5 \times 10^{37}$ ergs s$^{-1}$, we obtain a power law index of $-0.5 
\pm 0.1$, while for the higher luminosities the power law index is 
$-1.2^{+0.2}_{-0.4}$. There is again some flattening of the cumulative 
distribution towards the faint source end (Fig. \ref{lum_distr}{\em b}), 
probably caused by the incompleteness of our sample. These results for 
the M31 Core field are in general agreement with previous studies of the 
central part of M31 (\cite{Primini93,Supper01,Shirey01,Kong02}). 

\subsection{Comparison with previous results}

Studies of the luminosity distributions began as soon as individual 
X-ray sources in nearby galaxies were resolved (see Fabbiano 1995 for 
a review of early results). More sensitive surveys with {\it ROSAT} and
{\it Chandra} significantly increased the number of sources detected (see 
Read, Ponman \& Strickland (1997) for {\it ROSAT} and Fabbiano (2001) for 
{\it Chandra} review). Most previous studies, however, differ from our 
present work in two important aspects: (i) for most external galaxies, 
sensitivity limits were above or around 10$^{37}$ erg/s, while in our 
case we are able to study in detail population of sources between 
5$\times$10$^{35}$ and 10$^{37}$ erg/s; (ii) previous studies of luminosity 
distributions in M31 (\cite{TF91,Kong02}) were restricted to the 
central $17\arcmin - 20\arcmin$. Since the inner region of M31 ($r < 
15\arcmin$) is dominated by the bulge and spheroid populations, the disk 
population was not well sampled in previous surveys. Our results, however, 
can be directly compared to the luminosity distributions for Galactic sources
obtained by Grimm et al. (2001). The comparison shows that the luminosity 
function which we obtain for the core region is close to the luminosity 
function of LMXBs in our Galaxy: both distributions have a characteristic 
``knee'' at a few$\times 10^{37}$ ergs s$^{-1}$ and can be described by a 
broken power law or a cut-off power law models. The luminosity distribution 
for the two northern fields of M31 resembles the luminosity function for 
Galactic HMXBs: both have a power law-like form with a slope steeper than 
that of the corresponding M31 bulge/Galactic LMXB luminosity distributions; 
the M31 disk/Galactic HMXB distributions terminate at much lower luminosities 
($\sim 10^{37}$ ergs s$^{-1}$) than M31 bulge/Galactic LMXB luminosity 
distributions ($ > 10^{38}$ ergs s$^{-1}$). We conclude therefore that the 
population of 
fainter X-ray sources we found in disk of M31 is probably dominated by the 
HMXBs, while the bulge region inside 15$\arcmin$ is dominated by LMXBs by 
analogy with the Milky Way. Our results confirm that, as in Milky Way and 
other nearby galaxies, the younger population of HMXBs in M31 is typically 
less luminous than the older populations of LMXBs and globular clusters (see 
also Helfand \& Moran 2001 for a discussion of HMXB luminosities). In our 
study we see no evidence that the brightest sources in M31 are associated 
with young populations; in fact, we see the opposite effect. This is in 
contrast to Soria $\&$ Wu (2002), who invoke a population of bright young 
binaries to explain a large number of high luminosity objects observed in 
the starburst galaxies. That effect can be probably attributed to the 
differences in the galaxy type and ranges of X-ray luminosities under 
study.   

\section{CONCLUSIONS}

We analyzed {\it XMM-Newton} data for the three M31 fields and detected 
a total of 230 point-like sources. We built luminosity distributions 
separately for the bulge (inner 15$\arcmin$) and disk regions. A striking 
difference between the bulge and disk populations has been demonstrated for 
the first time. We report on the discovery of lower luminosity population 
in the disk of M31. Comparison of our results with the luminosity function 
obtained by Grimm et al. (2001) for the Milky Way allows us to conclude that 
lower-luminosity population in the disk is probably dominated by HMXBs, and 
that the luminosity distributions of X-ray sources in the bulge and disk of 
M31 are similar to the corresponding parts of our Galaxy.

\section{ACKNOWLEDGEMENTS}

We have used data obtained with {\em XMM-Newton} satellite. {\em XMM-Newton} 
is an ESA science mission with instruments and contributions directly funded 
by ESA Member States and the USA (NASA). We are grateful to the personnel of 
the {\em XMM-Newton} Science Operations Centre at VILSPA, Spain for satellite 
operations and expedited data preparation for scientific analysis. We are 
thankful to our colleagues from PV data analysis team for a fruitful 
collaboration in earlier analysis of M31 PV observations 
(\cite{Shirey01,Osborne01}). We are especially indebted to M.Watson, PI of 
M31 PV program.

\begin{table}
\small
\caption{XMM-Newton observations of M31 used in this analysis. 
\label{obslog}}
\begin{tabular}{cccccccc}
\hline
\hline
Date, UT & $T_{\rm start}$, UT & Field & Obs. ID  & RA (J2000)$^{a}$ & Dec (J2000)$^{a}$ & Exp.(MOS)$^{b}$ & Exp.(PN)$^{b}$\\
 &(h:m:s)&&&  (h:m:s)   &(d:m:s)&(s)&(s)\\             
\hline
25/06/2000 &10:44:42&M31 Core  &0112570401&00:42:43.0&41:15:46.1&34835&31021\\
05/01/2002 &06:28:31&M31 North1&0109270701&00:44:01.0&41:35:57.0&56810&49568\\
26/01/2002 &16:51:03&M31 North2&0109270301&00:45:20.0&41:56:09.0&32193&27982\\
\hline
\end{tabular}
\begin{list}{}{}
\item[$^{a}$] -- coordinates of the center of the field of view
\item[$^{b}$] -- instrument exposure used in the analysis 
\end{list}
\end{table}

\clearpage

\clearpage

\end{document}